\documentclass[twocolumn]{aastex62}

\graphicspath{{./}{figures/}}

\accepted{30 Decmeber 2018}

\shorttitle{Sub-Chandrasekhar Explosions}
\shortauthors{Polin et al.}

\begin{document}

\title{Observational Predictions for Sub-Chandrasekhar Mass Explosions: Further Evidence for Multiple Progenitor Systems for Type Ia Supernovae}

\correspondingauthor{Abigail Polin}
\email{abigail@berkeley.edu}

\author{Abigail Polin}
\affiliation{Department of Astronomy, University of California, Berkeley}
\affiliation{Lawrence Berkeley National Lab}

\author{Peter Nugent}
\affiliation{Department of Astronomy, University of California, Berkeley}
\affiliation{Lawrence Berkeley National Lab}

\author{Daniel Kasen}
\affiliation{Department of Astronomy, University of California, Berkeley}
\affiliation{Lawrence Berkeley National Lab}

\begin{abstract}
We present a numerical parameter survey of sub-Chandrasekhar mass white dwarf (WD) explosions. Carbon-oxygen WDs accreting a helium shell have the potential to explode in the sub-Chandrasekhar mass regime. Previous studies have shown how the ignition of a helium shell can either directly ignite the WD at the core-shell interface or propagate a shock wave into the the core causing a central ignition. We examine the explosions of WDs from 0.6 - 1.2 M$_\sun$  with helium shells of 0.01, 0.05 and 0.08 M$_\sun$. Distinct observational signatures of sub-Chandrasekhar mass WD explosions are predicted for two categories of shell size. Thicker-shell models show an early time flux excess, which is caused by the presence of radioactive material in the ashes of the helium shell, and red colors due to these ashes creating significant line blanketing in the UV through the blue portion of the spectrum. Thin shell models reproduce several typical Type Ia supernova signatures. We identify a relationship between Si II velocity and luminosity which, for the first time, identifies a sub-class of observed supernovae that are consistent with these models. This sub-class is further delineated by the absence of carbon in their atmospheres. We suggest that the proposed difference in the ratio of selective to total extinction between the high velocity and normal velocity Type Ia supernovae is not due to differences in the properties of the dust around these events, but is rather an artifact of applying a single extinction correction to two intrinsically different populations of supernovae.
\end{abstract}


\section{Introduction}
\label{sec:intro}
Sub-Chandrasekhar mass carbon-oxygen (CO) white dwarfs (WDs) have been discussed extensively as a possible progenitor for Type Ia supernovae (SNe~Ia). Early studies modeled the explosion of sub-Chandrasekhar mass CO WDs with a thick shell of accreted helium and found that a detonation in the helium shell can trigger an explosion of the CO core in what was called the ``double-detonation model" \citep{Woosley&Weaver94,Nomoto1982a,Nomoto1982b, Livne1990}.  However, at the time it was generally thought that these scenarios were not good candidates for Type Ia supernovae. The thick helium shells modeled produce too much $^{56}$Ni during nuclear burning for this to be a viable progenitor \citep{Hoeflich1996, Nugent1997}. The ashes of thick helium shells have also been shown to cause Fe-group line blanketing in the spectra, resulting in peak spectra redder than typical SNe~Ia \citep{Nugent1997,Kromer2010,Woosley&Kasen11}.  However, modern computing has allowed for more detailed studies of the nuclear reactions involved in these explosions, revealing that the minimal mass of a helium shell required to trigger an explosion via a double detonation is much smaller than those used in the early models \citep{Bildsten2007,Shen14,Fink2007, Fink2010}. Furthermore, recent Gaia observations of hypervelocity WDs have been proposed as possible surviving degenerate companions of double detonation explosions \citep{ShenD6}. The possibility and consequences of such thin helium shell explosions motivate the need to re-examine these progenitors for SNe~Ia. 

The upcoming era of all-sky, high cadence transient searches also motivates the consideration of potential rare transients. Sub-Chandrasekhar models with thick helium shells may still occur in nature even though they have yet to be detected in the current catalog of SNe~Ia, and thus are also worth considering.

Previous studies in both two \citep{Fink2007,Fink2010, Moll&Woosley13} and three dimensions \citep{Moll&Woosley13,Garcia-Senz} have shown that double detonation ignitions still occur robustly when multidimensional effects and multiple helium ignitions are considered. Determining the precise values of nucleosynthetic yields will necessitate multi-dimensional simulations with large nuclear networks. However, an advantage of 1-D simulations is that they enable large parameter surveys which can map out the landscape of all possible sub-Chandrasekhar mass explosions and elucidate clear qualitative trends.

This paper presents a survey of sub-Chandrasekhar mass explosion models and explores their observational relevance to SNe~Ia-like transients. The numerical setup used to simulate the explosion from the helium detonation through homology, and the radiative transport methods used to simulate observables are described in section \ref{sec:methods}. The hydrodynamic outflow and elemental yields synthesized from nuclear burning are discussed in section \ref{sec:nuc}. Light curves and spectra are examined in the context of observational predictions in sections \ref{sec:lightcurves} and \ref{sec:spectra} respectively. The results are summarized and future work is discussed in section \ref{sec:discussion}.

\section{Numerical Methods}
\label{sec:methods}
We model the explosion in several stages. First, the initial model is built by starting with a bare, isothermal WD progenitor. Next, a helium shell of some mass is added iteratively to the outermost layers of the WD, modelling the accretion of this mass slowly from a companion. During each step of this process the new progenitor is evolved to hydrostatic equilibrium to determine a stable central density of the WD. After the shell reaches a desired mass and equilibrium is achieved, burning is initiated within the helium layer. The resulting thermonuclear hydrodynamics are then evolved in 1-D. Each model is evolved until the ejecta reaches homologous expansion, after which we perform radiative transport calculations to produce light curves and spectra for each model.

\subsection{Initial Models and Parameter Space}
\label{subsec:initialmodels}

The initial models were created by the process described in \cite{Zingale2013}. Each model begins with an isothermal WD of a desired mass which is 50\% carbon and 50\% oxygen. The WD profile is then augmented with an isentropic helium shell on its surface. We use MAESTRO, a low Mach number hydrodynamics code \citep{1742-6596-1031-1-012024,maestro}, to build the model profile and ensure it begins in hydrostatic equilibrium. This checks that our choice of central density of the WD is appropriate to support the additional mass of the helium shell.
In the process of finding an equilibrium solution, a mixed transition layer is introduced between the CO WD and the helium layer. The purpose of this mixed region is to ensure there is no discontinuous change in composition. For most models this mixed region is resolved over 128 grid cells at the coarsest level of refinement. When the width of this layer is increased the nucleosynthic yields from the core of the white dwarf show no difference and the yields from the helium shell are similar to igniting the helium at an equivalently larger radius above the white dwarf (see table \ref{table:nucdata}).

This parameter survey varies the mass of the white dwarf ($M_{WD}$), the mass of the helium shell ($M_{He}$), the amount of material confined in the mixed region between the WD and the helium shell ($\delta$), and the location of the initial helium ignition ($R_{ign}$). The WD mass ranges from 0.6-1.2 M$_\sun$, in increments of 0.1 M$_\sun$, and the helium shell mass takes on the values 0.08, 0.05 and 0.01 M$_\sun$. Additional WD/He mass combinations of interest were explored less systematically. See table \ref{table:nucdata} for the comprehensive set of the 47 models explored in this survey as well as their conditions at the location of ignition.

\subsection{Hydrodynamics and Nuclear Processes}
\label{subsec:castro}

The compressible Eulerian hydrodynamics code Castro is used to follow the progression of the explosions. Castro allows for adaptive mesh refinement (AMR) which enables the selective increase of resolution in areas of interest in the simulation; specifically, in areas where nuclear burning is active \citep{1742-6596-1031-1-012024,CASTRO}. Each simulation is allowed 4 levels of refinement, resulting in a minimal resolution of order $10^3$ cm. 

We enable Castro's built-in monopole gravity and the Helmholtz equation of state \citep{Timmes00}. A 13 isotope $\alpha$-chain nuclear network is used to monitor the nuclear reactions \citep{Timmes99}. The network includes ($\alpha$,$\gamma$) and ($\gamma$, $\alpha$) reactions for 13-isotopes: $^{4}$He, $^{12}$C, $^{16}$O, $^{20}$Ne, $^{24}$Mg, $^{28}$Si, $^{32}$S, $^{36}$Ar, $^{40}$Ca, $^{44}$Ti, $^{48}$Cr, $^{52}$Fe, and $^{56}$Ni. For temperatures above 2.5$\times$10$^9$K, ($\alpha$,p)(p,$\gamma$) reactions (and their inverse reactions) are included for 8 additional isotopes: $^{27}$Al, $^{31}$P, $^{35}$Cl, $^{39}$K, $^{43}$Sc, $^{47}$V, $^{51}$Mn, and $^{55}$Co. 

Several additional calculations were performed with a larger reaction network consisting of 21 isotopes. The 21 isotope network includes all of the reactions in the 13 isotope network (including those additional reactions for temperatures above 2.5$\times$10$^9$K) as well as $^{1}$H, $^{3}$He, $^{14}$N, $^{56}$Cr, $^{54}$Fe, and $^{56}$Fe. These additional isotopes allow for the distinction of iron group elements toward the end of the $\alpha$-chain in more detail, with the presence of three more radioactive isotopes. In the examples considered, the larger network did not yield significantly different results and it was determined that the 13 isotope network was sufficient for the scope of this study.

\subsection{Ignitions and Explosion Mechanisms}
\label{subsec:ignitions}

After an initial model is shown to have reached an equilibrium state it is imported into Castro where the remaining hydrodynamics calculations are performed. A radius is chosen at which to trigger an ignition in the helium layer by heating a single grid cell in the coarsest level of refinement (order $10^5$ cm). Once nucleosynthesis is initiated, all artificial heating is turned off and the hydrodynamics are evolved through homology.

There are two promising paths to explosion for these sub-Chandrasekhar mass WDs. In the double detonation scenario, the ignition of the helium shell sends a shock wave into the CO WD igniting the WD at the center of its core \citep{Woosley&Weaver94,Nomoto1982a,Nomoto1982b}. The second scenario is known as the direct-dive or edge-lit detonation, where the ignition of the helium shell directly ignites the CO WD at its interface with the helium shell \citep{Shen14,Moll&Woosley13}. Progenitors modeled in this study exhibit both ignition mechanisms depending on the properties (density, temperature and pressure) of the ignition site within the helium shell.

Models are initially ignited at the base of the convective region in the helium shell. This is a natural choice because it is likely where the helium reaches the highest temperature and pressure while still under hydrostatic equilibrium. Both edge-lit explosions and traditional double detonations occur. It has been shown that an edge-lit explosion modeled in 1-D are qualitatively similar to a double detonation of the same progenitor \citep{Woosley&Kasen11} but it is expected that this degeneracy will break down in higher dimensional studies. Given the 1-D nature of this study, we choose to focus on the double detonations, which are likely to have fewer asymmetries moving into higher dimensions than their edge-lit counterparts.

After examining the initial parameter space of models, further choices for ignition location were explored. It is possible to ignite the edge-lit models as double detonations, by igniting the helium at a location of lower temperature, pressure and density (or a larger radius) within the shell. After examining all of the results where models were ignited at the interface between the WD and the shell, each of the initial models that led to an edge-lit explosion was re-ignited at a minimal radius required to produce a double detonation explosion (see table \ref{table:nucdata}).

\subsection{Radiative Transport Methods}
\label{subsec:sedona}

After the SN ejecta reaches homologous expansion we use the Sedona code \citep{sedona} to create synthetic light curves and spectra for each model. Sedona is a multi-dimensional time dependent radiation transport code which uses Monte Carlo methods to propagate photons. The calculations are performed under the assumption of local thermal equilibrium (LTE) to determine ionization and excitation fractions in the ejecta. Energy is generated through three radioactive decay chains: $^{56}$Ni$\rightarrow^{56}$Co$\rightarrow^{56}$Fe, $^{48}$Cr$\rightarrow^{48}$V$\rightarrow^{48}$Ti and $^{52}$Fe$\rightarrow^{52}$Mn$\rightarrow^{52}$Cr.

\section{Nucleosynthetic Yields}
\label{sec:nuc}

\begin{figure}[tph!]
  \centering
  \includegraphics[width=\columnwidth]{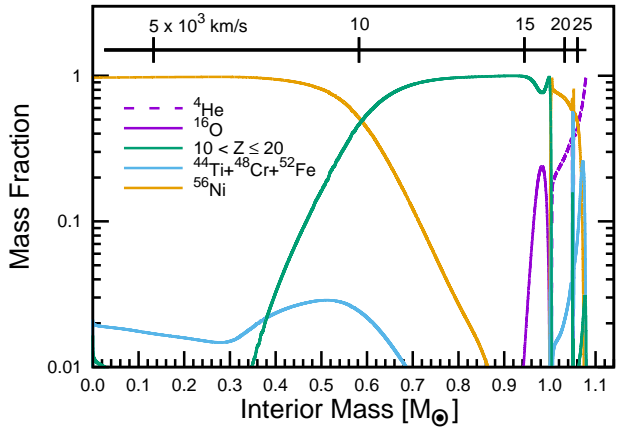}
  \caption{Elemental composition of the supernova ejecta as a function of interior mass for a 1.0 M$_\sun$  WD with a 0.08 M$_\sun$  helium shell that has reached homology. The mass fractions $^{4}$He (dashed) and $^{16}$O (solid) are shown in purple. Summed mass fractions of intermediate mass elements (IMEs) ($10<Z<22$) in green, radioactive material of $  22 \leq Z < 28$ in blue, and $^{56}$Ni is shown in yellow. $^{12}$C is almost completely burnt in this model with a mass fraction of never reaching above $10^{-4}$. The bar on top of the figure indicates the ejecta velocity in increments of 5000 km/s. Everything for an interior mass greater than 1.0 M$_\sun$  is synthesized from the burnt helium shell and everything less than 1.0 M$_\sun$  is from the underlying WD. This is one of the heaviest shells modelled in this paper, so a large amount of radioactive material is produced from the shell.}
  \label{fig:composition}
\end{figure}

This section examines the nucleosynthesis yields of the explosion models.  First, the overall composition of the material produced is investigated.  Next, the composition of the fastest ejecta is examined. This outermost material, which is created from the burning of the helium shell, dominates the early lightcurve and spectral features, imprinting a characteristic signature on the observed transient. Underneath the ashes of the helium shell the ejecta looks more like a standard thermonuclear WD explosion. The $\alpha$-chain burning becomes more complete when burning occurs at higher density, temperature and pressure. Thus, the center of the ejecta is composed of mainly heavy, radioactive material, while intermediate mass and lighter elements are present further out (see Figure~\ref{fig:composition} for an example from a thick helium shell model).

\subsection{Overall Nucleosynthetic Yields}
\label{subsec:totalmaterial}

\begin{figure}[tbph]
  \centering
  \includegraphics[width=\columnwidth]{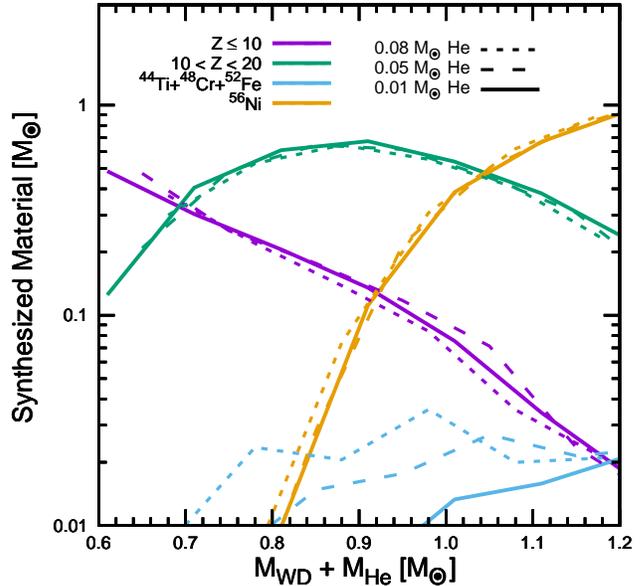}
  \caption{Nucleosynthesis yields for models ignited at the interface between the helium shell and the CO WD. Models with the same mass helium shell are connected with lines of the same dash type (solid line for 0.01 M$_\sun$ He, long dashes for 0.05 M$_\sun$ He, and short dashes for 0.08 M$_\sun$ He). Summed masses of elements produced with $Z \leq 10$ (mostly $^{16}$O) are shown in purple, IMEs ($10<Z<22$) in green, radioactive material of $  22 \leq Z < 28$ in blue, and $^{56}$Ni is shown in yellow. As the central density of the progenitor increases, nuclear burning completes more of the $\alpha$-chain network producing heavier elements. Models whose total mass exceeds 0.9 M$_\sun$  produce $^{56}$Ni in quantities required to power normal SN~Ia-like lightcurves.}
  \label{fig:totalmaterial}
\end{figure}

\begin{figure}[tpbh!]
  \centering
  \includegraphics[width=\columnwidth]{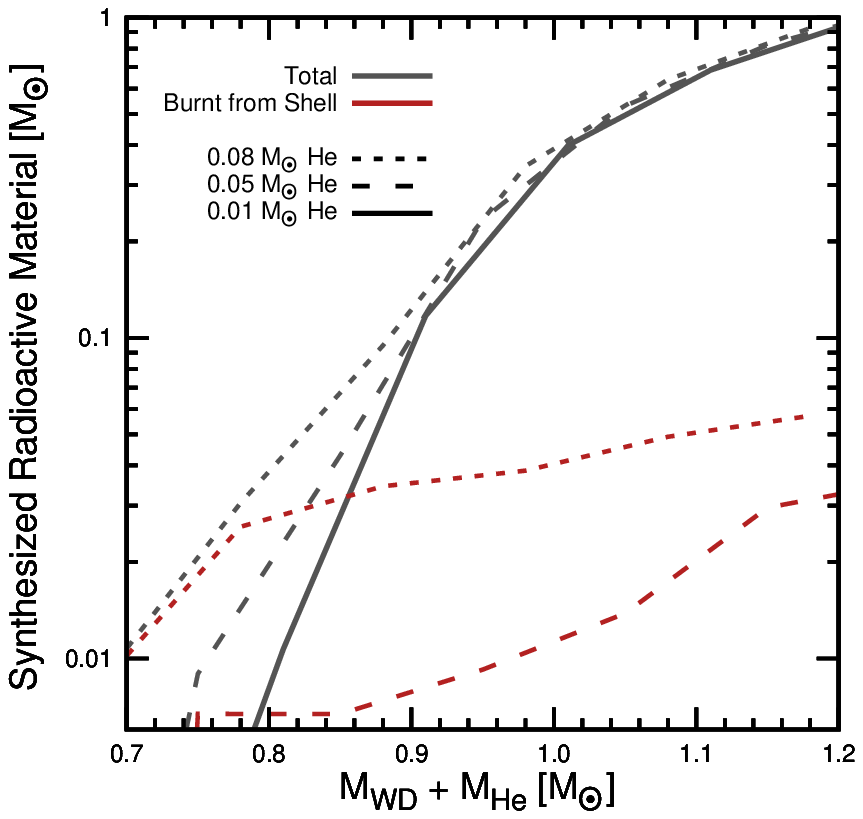}
  \caption{The total radioactive material produced (grey) compared with the summed mass of radioactive material burnt from the helium shell (red). Solid lines are models with 0.01 M$_\sun$ He, long dashes have 0.05 M$_\sun$ He, and short dashes 0.08 M$_\sun$ He. Models with 0.01 M$_\sun$  helium shells produce less than 0.003 M$_\sun$  of radioactive material in their outermost ejecta and are not plotted. For models with a total mass less than 0.9 M$_\sun$  the amount of radioactive material produced in the outermost ejecta is comparable to the total radioactive material produced. Observable signatures of this heavy, radioactive material present in the outermost ejecta are discussed in section~\ref{sec:lightcurves}.}
  \label{fig:shellmaterial}
\end{figure}

Figure~\ref{fig:totalmaterial} shows the total abundances produced by models ignited at the interface between the helium shell and the CO WD. As the total mass of the initial model increases, so does its central density, temperature and pressure. As burning occurs the more massive models burn more completely through the $\alpha$-chain network, and heavier elements are created. Specifically the amount of $^{56}$Ni produced is significantly larger for models with a total mass above 0.9 M$_\sun$, reaching the quantities required to power SN~Ia-like brightness through radioactive decay.

These results are in line with expectations from previous studies which have established a relationship between $^{56}$Ni and total mass of the system. We compare our $^{56}$Ni yields from core burning to the closest matched models (when comparing both the mass of the helium shell and the underlying WD) in four previous studies. The $^{56}$Ni yields in this study show close agreement to the amount of $^{56}$Ni produced as a function of total mass to the bare sub-Chandrasekhar mass WDs (those with no helium shell) in \cite{Sim2010}. However, more recent studies of these bare WDs in \cite{Shen2018} show significantly more $^{56}$Ni than produced in our study for the lower-mass models. The bare 0.9 M$_\sun$  model produced a factor of 2 more $^{56}$Ni than the 0.9 M$_\sun$  WD with a 0.01 M$_\sun$  helium shell in this study. Our models also show similar yields to the traditional 2-dimensional double detonation explosions modeled in \cite{Fink2010} and the 3-dimensional rotating WDs modeled in \cite{Garcia-Senz} when examining the amount of $^{56}$Ni produced from the underlying WD. For example the \cite{Fink2010} 0.92 M$_\sun$  WD with a 0.084 M$_\sun$  helium shell produces 0.34 M$_\sun$  $^{56}$Ni in the core, and the 0.9 M$_\sun$  WD with a 0.08 M$_\sun$  helium shell in this study produces 0.31 M$_\sun$  $^{56}$Ni in the core, and our 1.1 M$_\sun$ WD with a 0.05 M$_\sun$ model produces only 10\% more $^{56}$Ni than the \cite{Garcia-Senz} 1.1M$_\sun$ WD with 0.052 M$_\sun$ helium.

\begin{deluxetable*}{cccccccccccccccccc}
\tablecaption{Initial Model Parameters and Nucleosynthetic Yields. \label{table:nucdata}}
\tabletypesize{\scriptsize}
\setlength{\tabcolsep}{4pt}
\tablehead{
\colhead{$M_{WD}$} & \colhead{$M_{He}$} & \colhead{type} & \colhead{$\delta$} & \colhead{$\rho_{c}$} & \colhead{$R_{ign}$} & \colhead{$P_{ign}$} & \colhead{$\rho_{ign}$} & \colhead{$M_{tot}$} & \colhead{$M_{tot}$} & \colhead{$M_{tot}$} & \colhead{$M_{tot}$} & \colhead{$M_{shell}$} & \colhead{$M_{shell}$} & \colhead{$M_{shell}$} & \colhead{$M_{shell}$} \\
\colhead{}&\colhead{}&\colhead{}&\colhead{}&\colhead{}&\colhead{}&\colhead{}&\colhead{}& \colhead{$ Z \leq 10$} & \colhead{IME} & \colhead{$Z\geq22$} &\colhead{$ ^{56}Ni$} & \colhead{$ Z \leq 10$} & \colhead{IME} & \colhead{$Z\geq22$} &\colhead{$ ^{56}Ni$} \\
\colhead{[M$_{\sun}$]} & \colhead{[M$_\sun$]} & \colhead{}& \colhead{[$cm$]} & \colhead{[$g/cm^3$]} &\colhead{[$cm$]}&\colhead{[$dyne/cm^2$]}&\colhead{[$g/cm^3$]}&\colhead{[M$_\sun$]}&\colhead{[M$_\sun$]}&\colhead{[M$_\sun$]}&\colhead{[M$_\sun$]}&\colhead{[M$_\sun$]}&\colhead{[M$_\sun$]}&\colhead{[M$_\sun$]}&\colhead{[M$_\sun$]} \\
} 
\startdata
0.60&0.010&$\bigcirc$&5e06&3.8e06&7.7e08&2.0e20&1.5e04&4.8e-01&1.2e-01&4.5e-07&7.2e-08&9.8e-03&3.8e-04&5.1e-12&1.3e-12\\
0.70&0.010&$\bigcirc$&5e06&6.5e06&7.0e08&3.5e20&2.7e04&3.0e-01&4.0e-01&6.5e-04&5.5e-04&8.9e-03&1.0e-03&9.5e-09&1.2e-12\\
0.80&0.010&$\bigcirc$&5e06&1.1e07&6.3e08&6.1e20&4.4e04&2.1e-01&6.1e-01&1.1e-02&9.8e-03&8.0e-03&1.9e-03&5.5e-05&3.1e-12\\
0.85&0.010&$\bigcirc$&5e06&1.5e07&6.0e08&8.0e20&5.6e04&1.5e-01&6.7e-01&3.6e-02&3.3e-02&7.2e-03&2.5e-03&2.7e-04&5.2e-11\\
0.90&0.010&$\bigcirc$&5e06&2.0e07&5.6e08&1.0e21&7.2e04&1.4e-01&6.7e-01&1.2e-01&1.1e-01&6.2e-03&3.1e-03&6.6e-04&6.2e-09\\
1.00&0.010&$\bigcirc$&5e06&3.6e07&5.0e08&1.7e21&1.1e05&7.5e-02&5.4e-01&4.0e-01&3.9e-01&6.1e-03&3.5e-03&3.7e-04&1.1e-07\\
1.10&0.010&$\bigcirc$&5e06&7.3e07&4.3e08&3.0e21&1.8e05&3.4e-02&3.8e-01&6.9e-01&6.7e-01&6.0e-03&3.5e-03&5.1e-05&1.3e-07\\
1.20&0.010&$\bigcirc$&5e06&1.8e08&3.5e08&5.7e21&2.9e05&1.7e-02&2.3e-01&9.6e-01&9.4e-01&2.0e-03&4.1e-03&3.8e-03&2.8e-05\\
\hline
1.00&0.020&$\bigcirc$&5e06&3.8e07&4.7e08&4.5e21&2.4e05&7.0e-02&5.3e-01&4.2e-01&4.1e-01&1.2e-02&1.7e-03&5.8e-03&1.5e-03 \\
0.85&0.025&$\bigcirc$&5e06&1.6e07&5.6e08&2.7e21&1.6e05&1.4e-01&6.7e-01&5.9e-02&5.2e-02&1.9e-02&1.7e-03&3.9e-03&1.7e-05 \\
1.00&0.030&$\bigcirc$&5e06&4.1e07&4.5e08&8.3e21&3.8e05&6.6e-02&5.1e-01&4.6e-01&4.4e-01&1.7e-02&1.9e-03&1.1e-02&6.4e-03 \\
1.00&0.040&$\bigcirc$&5e06&4.4e07&4.3e08&1.3e22&5.3e05&6.2e-02&4.9e-01&4.9e-01&4.7e-01&2.0e-02&2.4e-03&1.8e-02&1.2e-02 \\
0.95&0.045&$\bigcirc$&5e06&3.2e07&4.6e08&1.2e22&4.9e05&8.0e-02&5.6e-01&3.6e-01&3.4e-01&2.4e-02&1.6e-03&1.9e-02&1.2e-02 \\
\hline
0.60&0.050&$\bigcirc$&5e06&4.5e06&6.6e08&1.9e21&1.2e05&4.7e-01&2.1e-01&6.7e-05&2.3e-05&5.0e-02&7.1e-06&2.0e-11&5.0e-12 \\
0.70&0.050&$\bigcirc$&5e06&7.8e06&6.0e08&3.3e21&1.9e05&2.5e-01&4.9e-01&9.0e-03&2.0e-03&3.6e-02&6.7e-03&6.7e-03&3.1e-06 \\
0.80&0.050&$\bigcirc$&5e06&1.4e07&5.4e08&5.9e21&3.0e05&1.8e-01&6.4e-01&4.3e-02&2.8e-02&4.1e-02&1.9e-03&6.7e-03&9.4e-05 \\
0.90&0.050&$\bigcirc$&5e06&2.5e07&4.8e08&1.1e22&4.6e05&1.2e-01&6.2e-01&2.3e-01&2.2e-01&4.0e-02&9.9e-04&9.2e-03&2.5e-03 \\
0.92&0.050&$\bigcirc$&5e06&2.8e07&4.7e08&1.2e22&4.9e05&9.1e-02&5.9e-01&2.9e-01&2.7e-01&2.7e-02&1.6e-03&2.1e-02&1.2e-02 \\
0.94&0.050&$\bigcirc$&5e06&3.1e07&4.6e08&1.3e22&5.2e05&8.2e-02&5.6e-01&3.5e-01&3.3e-01&2.6e-02&2.5e-03&2.1e-02&1.3e-02 \\
0.95&0.050&$\bigcirc$&5e06&3.3e07&4.5e08&1.4e22&5.5e05&7.8e-02&5.4e-01&3.8e-01&3.6e-01&2.6e-02&1.9e-03&2.2e-02&1.4e-02 \\
0.96&0.050&$\bigcirc$&5e06&3.6e07&4.4e08&1.5e22&5.9e05&7.4e-02&5.3e-01&4.1e-01&3.9e-01&2.5e-02&1.4e-03&2.3e-02&1.5e-02 \\
1.00&0.050&$\bigcirc$&5e06&4.7e07&4.2e08&1.9e22&6.9e05&6.7e-02&4.5e-01&5.3e-01&5.0e-01&2.6e-02&8.0e-03&1.6e-02&2.3e-03 \\
1.00&0.050&$\bigcirc$&1e07&4.7e07&4.2e08&1.9e22&6.9e05&4.9e-02&4.7e-01&5.3e-01&5.1e-01&2.1e-02&3.7e-03&2.5e-02&1.9e-02 \\
1.00&0.050&$\bigcirc$&5e06&4.7e07&4.6e08&5.3e21&3.1e05&5.0e-02&4.7e-01&5.3e-01&5.1e-01&2.1e-02&1.9e-03&2.6e-02&2.2e-02 \\
1.10&0.050&$\bigcirc$&5e06&1.0e08&3.7e08&1.9e22&7.4e05&2.4e-02&3.1e-01&8.2e-01&8.0e-01&1.9e-02&1.7e-03&3.0e-02&2.6e-02 \\
1.10&0.050&$\bigcirc$&5e06&1.0e08&3.8e08&1.3e22&5.7e05&1.9e-02&3.1e-01&8.3e-01&8.1e-01&1.7e-02&1.4e-03&3.1e-02&3.0e-02 \\
1.10&0.050&$\bigstar$&5e06&1.0e08&3.5e08&3.8e22&1.1e06&6.4e-02&2.4e-01&8.5e-01&8.3e-01&1.9e-02&1.9e-03&2.9e-02&2.1e-02 \\
1.20&0.050&$\bigcirc$&5e06&2.7e08&3.1e08&1.9e22&7.7e05&1.3e-02&1.5e-01&1.1e00&1.1e00&1.3e-02&8.7e-04&3.6e-02&3.5e-02 \\
1.20&0.050&$\bigstar$&5e06&2.7e08&2.8e08&9.3e22&2.1e06&1.4e-02&1.3e-01&1.1e00&1.1e00&1.4e-02&9.0e-04&3.5e-02&3.3e-02 \\
\hline
0.95&0.055&$\bigcirc$&5e06&3.4e07&4.4e08&1.6e22&6.3e05&7.7e-02&5.3e-01&4.0e-01&3.7e-01&2.7e-02&2.1e-03&2.6e-02&1.2e-02 \\
1.00&0.060&$\bigcirc$&5e06&5.0e07&4.1e08&1.9e22&7.2e05&5.3e-02&4.4e-01&5.7e-01&5.4e-01&2.5e-02&2.8e-03&3.2e-02&1.7e-02 \\
1.00&0.070&$\bigcirc$&5e06&5.3e07&4.1e08&1.9e22&7.4e05&4.3e-02&4.2e-01&6.0e-01&5.8e-01&2.7e-02&2.7e-03&4.0e-02&3.4e-02 \\
\hline
0.60&0.080&$\bigcirc$&5e06&5.2e06&6.1e08&4.2e21&2.3e05&3.7e-01&3.0e-01&8.3e-03&1.4e-04&5.9e-02&1.2e-02&8.1e-03&1.2e-07 \\
0.70&0.080&$\bigcirc$&5e06&9.1e06&5.5e08&7.5e21&3.6e05&2.2e-01&5.3e-01&3.1e-02&7.1e-03&4.8e-02&6.3e-03&2.6e-02&2.7e-03 \\
0.80&0.080&$\bigcirc$&5e06&1.6e07&4.9e08&1.3e22&5.4e05&1.4e-01&6.5e-01&9.5e-02&7.4e-02&4.1e-02&4.0e-03&3.4e-02&1.7e-02 \\
0.90&0.080&$\bigcirc$&5e06&2.9e07&4.5e08&1.9e22&7.2e05&8.4e-02&5.5e-01&3.4e-01&3.1e-01&3.6e-02&5.3e-03&3.9e-02&1.3e-02 \\
0.90&0.080&$\bigstar$&5e06&2.9e07&4.4e08&2.4e22&8.1e05&1.6e-01&3.9e-01&4.3e-01&4.1e-01&3.3e-02&3.0e-03&4.3e-02&3.0e-02 \\
1.00&0.080&$\bigcirc$&5e06&5.7e07&4.1e08&1.9e22&7.5e05&3.6e-02&4.0e-01&6.4e-01&6.2e-01&2.9e-02&1.7e-03&4.9e-02&4.4e-02 \\
1.00&0.080&$\bigstar$&5e06&5.7e07&3.8e08&4.6e22&1.3e06&9.2e-02&3.0e-01&6.9e-01&6.7e-01&2.9e-02&1.9e-03&4.8e-02&3.8e-02 \\
1.10&0.080&$\bigcirc$&5e06&1.3e08&3.5e08&1.9e22&7.7e05&2.1e-02&2.3e-01&9.3e-01&9.1e-01&2.1e-02&1.4e-03&5.7e-02&5.5e-02 \\
1.10&0.080&$\bigstar$&5e06&1.3e08&3.1e08&9.6e22&2.1e06&2.4e-02&1.9e-01&9.6e-01&9.4e-01&2.2e-02&1.3e-03&5.6e-02&5.3e-02 \\
1.20&0.080&$\bigstar$&5e06&4.0e08&2.4e08&2.7e23&4.4e06&1.4e-02&6.3e-02&1.2e00&1.2e00&1.4e-02&6.3e-04&6.5e-02&6.4e-02 \\
\hline
1.00&0.090&$\bigcirc$&5e06&6.1e07&4.0e08&1.9e22&7.5e05&3.1e-02&3.8e-01&6.7e-01&6.5e-01&3.0e-02&2.0e-03&5.8e-02&5.3e-02 \\
0.70&0.100&$\bigcirc$&5e06&1.0e07&5.2e08&1.2e22&4.9e05&2.1e-01&5.5e-01&4.9e-02&2.1e-02&6.3e-02&3.6e-03&3.3e-02&9.2e-03 \\
1.00&0.100&$\bigcirc$&5e06&6.6e07&4.0e08&1.9e22&7.6e05&3.1e-02&3.6e-01&7.1e-01&7.0e-01&3.1e-02&9.5e-04&6.8e-02&6.3e-02 \\
1.10&0.100&$\bigcirc$&5e06&1.6e08&3.4e08&1.9e22&7.8e05&2.3e-02&1.5e-01&1.0e00&1.0e00&2.3e-02&7.3e-04&7.6e-02&7.4e-02 \\
1.20&0.100&$\bigcirc$&5e06&5.4e08&2.6e08&1.9e22&8.0e05&1.5e-02&2.3e-02&1.3e00&1.2e00&1.5e-02&1.0e-04&8.5e-02&8.4e-02
\enddata
\tablecomments{ \\ $R_{ign}$, $P_{ign}$ and $\rho_{ign}$ denote the conditions of the initial model at the location of the helium ignition.\\
A $\bigcirc$ indicates a double detonation while a $\bigstar$ indicates an edge-lit ignition.} 
\end{deluxetable*}

\subsection{Outermost ejecta: material produced from the helium shell}
\label{subsec:shellmaterial}

\begin{figure*}[!htb]
  \centering
  \includegraphics[width=1.0\columnwidth]{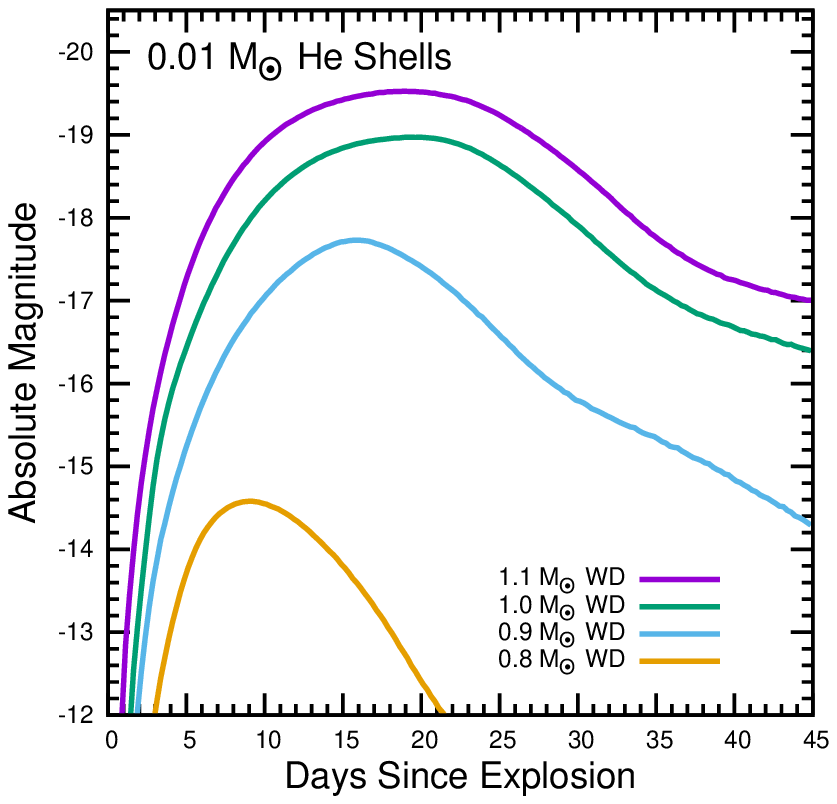}\hspace{-4em}
  \includegraphics[width=1.0\columnwidth]{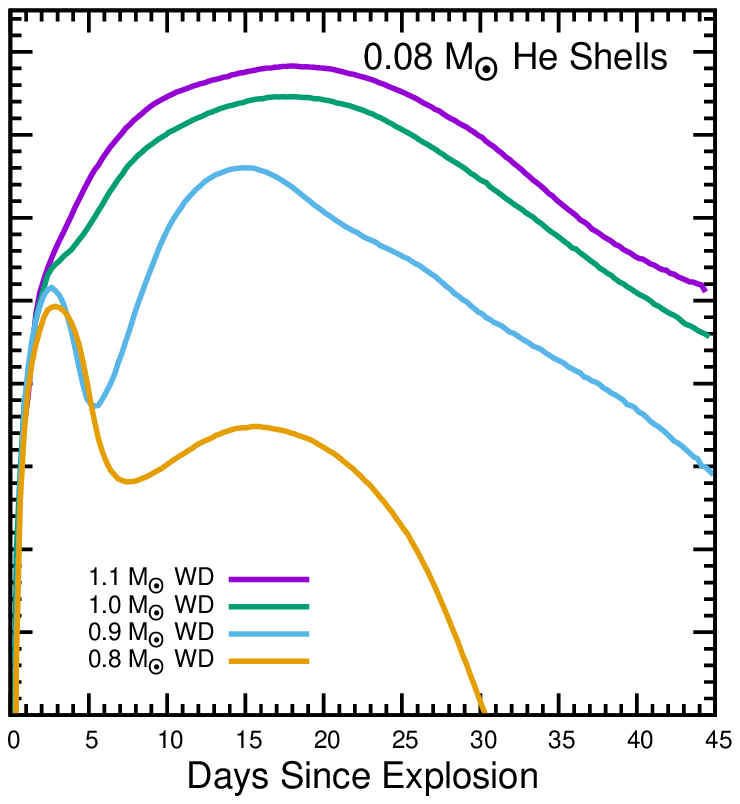}
\caption{g-band light curves of double detonation explosions for thin helium shells (0.01 M$_\sun$) on the left and thick helium shells (0.08 M$_\sun$) on the right. The figure shows a 1.1 M$_\sun$ WD in purple, a 1.0 M$_\sun$ WD in green, a 0.9 M$_\sun$ WD in blue and a 0.8 M$_\sun$ WD in orange. The thick helium shell models on right exhibit an early time flux excess due to radioactive material ($^{48}$Cr, $^{52}$Fe, and $^{56}$Ni) in the outermost ejecta, while the thin shell models on the left exhibit a smooth rise.}
\label{fig:gbandshells}
\end{figure*}

A distinct feature of a sub-Chandrasekhar mass WD explosion is the presence of heavy, radioactive material in the outermost ejecta, a result of the burning of the initial helium shell. Figure~\ref{fig:shellmaterial} shows the amount of radioactive material produced from the burnt helium shell compared with the overall radioactive material produced. The heavier the helium shell the more radioactive material is produced on the surface of the explosion. Thin helium shells do not produce a significant amount of radioactive material in the outermost ejecta.

As this material decays it creates an excess flux in the early observed lightcurves not present in traditional Chandrasekhar mass WD explosions. The presence of this material is the primary distinguishing factor between sub-Chandrasekhar and
Chandrasekhar mass explosions. Figure~\ref{fig:shellmaterial} shows that in models with a total mass less than 0.9 M$_\sun$  a significant fraction of the total radioactive material produced is synthesized in the helium shell. In section \ref{sec:lightcurves} we discuss the effect of this radioactive material on the lightcurves of the thick shell models. 

There is some discrepancy when comparing our yields from the burnt helium shell to those in the literature. The models in \cite{Fink2010} (2-D) and \cite{Garcia-Senz} (3-D) produce 10 times less $^{56}$Ni than our comparable helium shells. However, we show good agreement with the 3-D models of burning helium shells in \cite{Moll&Woosley13} which only produce 20\% less $^{56}$Ni in the shells than our models, so this discrepancy is not purely a multi-dimsenional effect. All of the compared models show close agreement in the summed masses of other radioactive material (Z$\geq$22 not including $^{56}$Ni) produced by the helium shells. The exact yields burnt from these shells remains an open question which should be examined in future multi-dimensional studies with careful consideration of the nucleosynthesis of helium and a realistic modelling of shell density profiles expected from accretion.

\section{Light Curves}
\label{sec:lightcurves}

The radiative transport calculations from Sedona provide synthetic lightcurves and spectra of each model. For the remainder of this paper, models will be separated into two categories: thin helium shells and thick helium shells. The former likely evolve from a degenerate helium WD binary companion \citep{Bildsten2007,Shen14} and look more like standard SNe~Ia. The later evolve from a non-degenerate helium star companion \citep{Woosley&Weaver94,Livne1990,Nomoto1982a}, exhibit more exotic observable signatures, and thus should be considered a sub-class of their own - with potentially only one of these events convincingly seen to date \citep{2017Natur.550...80J}.

\begin{figure}[tbph]
  \centering
  \includegraphics[width=\columnwidth]{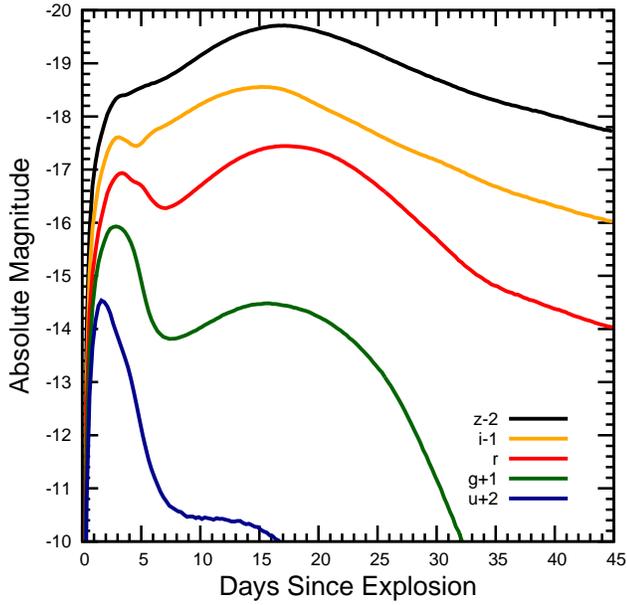}
\caption{ugriz colors for a thick helium shell model; a 0.8 M$_\sun$  WD with a 0.08 M$_\sun$  helium shell. For this model the radioactive material produced by the shell is comparable to the amount produced by the WD. The early excess flux created from the decay of the outermost radioactive material is similar in magnitude to the $^{56}$Ni powered peak. However the early flux excess is not uniform across the different bands.}
\label{fig:ugriz}
\end{figure}

Figure~\ref{fig:gbandshells} shows the g-band lightcurves for models with the extrema of helium shell masses explored in this survey: the very thick shells, 0.08 M$_\sun$, and the very thin shells, 0.01 M$_\sun$. The presence of radioactive material from the burning of the helium shell can create an excess in the flux at early times in the light curves of the models. The thin shells in Figure~\ref{fig:gbandshells} do not produce enough radioactive material from the shell burning to exhibit this early time excess, and the early rise looks like those typical of SNe~Ia. The thick shells, however, produce a large amount of heavy, radioactive elements having $Z \geq 22$ in the outermost ejecta, resulting in a significant excess in the first few days after explosion due to additional radioactive heating. The result is a bi-modal lightcurve with the first peak, a few days after explosion, powered primarily by radioactive material in the outermost ejecta, followed by a more standard $^{56}$Ni powered rise typical of thermonuclear SNe~Ia light curves. However the time of peak brightness will occur later than a pure $^{56}$Ni powered rise due to the decay of $^{48}$V (a daughter species of the $^{48}$Cr leftover from the helium shell) which has a half-life of 15.97 days.

For models with enough mass in the helium shell the early flux excess can be observed in surveys that explore the lightcurves of SNe 4 to 5 magnitudes below peak brightness. See table \ref{table:lcdata} for the magnitudes at both peak and early excess (when presented) for all models. When exhibited, this flux excess is present across all observable bands with the greatest differences occurring in the near-UV (See Figure~\ref{fig:ugriz}).

\begin{figure}[tbph]
  \centering
  \includegraphics[width=\columnwidth]{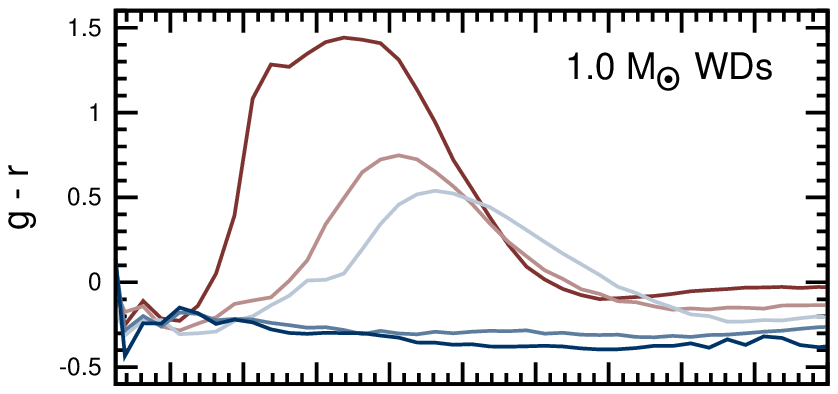} \vspace{-4em} \\
  \includegraphics[width=\columnwidth]{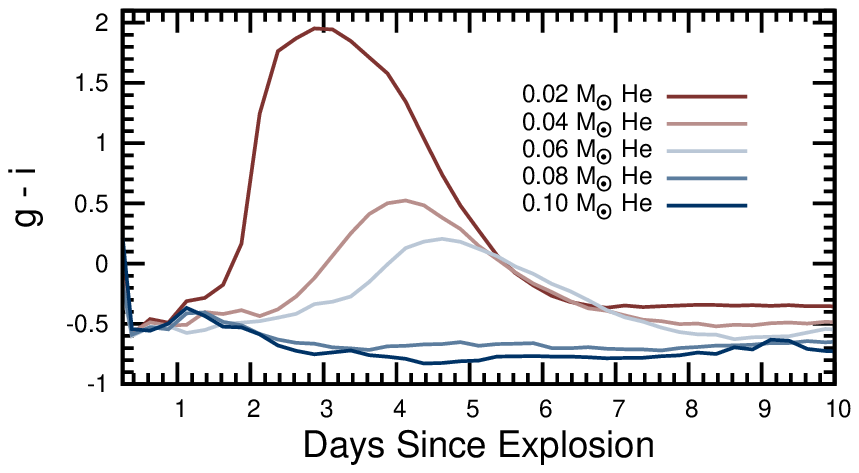}
\caption{g-r and g-i colors for a 1.0 M$_\sun$ WD with varying helium shell masses. It is interesting to note that models with less pronounced early excess peaks show the most significant reddening. This is due to the production of additional radioactive material by the heavier shell, on the same mass of underlying WD, heating the supernova at early times though radioactive decay.}
\label{fig:g-r}
\end{figure}

As a result the colors shown in Figure~\ref{fig:g-r} can become significantly red directly after the flux excess, before trending bluer again for the $^{56}$Ni powered rise. This figure depicts the $g-r$ and $g-i$ colors for 1.0 M$_\sun$ with several different mass helium shells. The early red evolution is a direct prediction that can identify a double detonation explosion, even when the helium shell is thin and the flux excess is not large. We note the change in behavior between the 0.06 and 0.08 M$_\sun$  helium shells on the 1.0 M$_\sun$  WD in Figure~\ref{fig:g-r}. This trend of a bluer feature for larger shells (on the same underlying WD) is caused by the additional heating from the radioactive material produced during helium burning. The 1.0 M$_\sun$  WDs shells above 0.06 M$_\sun$ produce enough radioactive material to cause significant heating at early time. Large shells on smaller WDs, however, can still appear quite red (see Figure~\ref{fig:ugriz}). This observational signature of a sub-Chandrasekhar mass Ia explosion can be detected photometrically though high cadence multi-band observations at very early times.

\begin{figure}[tbph]
  \centering
  \includegraphics[width=\columnwidth]{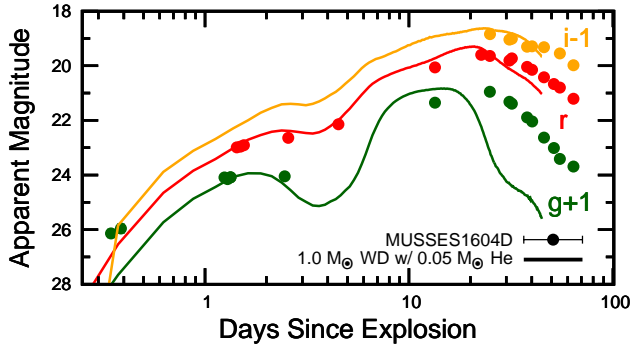}
\caption{SN 2016jhr (MUSSES1604D) \citep{Jiang2017} plotted against this survey's best fit model for the early flux excess, a 1.0 M$_\sun$  WD with a 0.05 M$_\sun$ helium shell. i-band (-1) is plotted in yellow, r-band in red, and g-band (+1) is green. While a double detonation explosion can reproduce the magnitudes at peak and during the early flux excess, the overall width is too narrow in all bands, most noticeably in g-band.}
\label{fig:musses}
\end{figure}

\begin{deluxetable}{ccccccccc}
\tablecaption{Observables from Synthetic Light Curves. \label{table:lcdata}}
\tabletypesize{\scriptsize}
\setlength{\tabcolsep}{2pt}
\tablehead{
\colhead{$M_{WD}$} & \colhead{$M_{He}$} & \colhead{type} & \colhead{V-mag} & \colhead{V-mag}& \colhead{B-mag} & \colhead{B-mag} & \colhead{$\Delta m_{15,V}$} & \colhead{B-V}\\
\colhead{} & \colhead{} & \colhead{} & \colhead{excess} & \colhead{peak}& \colhead{excess} & \colhead{peak} & \colhead{} & \colhead{}
}
\startdata
0.70&0.010&$\bigcirc$& &-11.58& &-10.36&3.20&1.22\\
0.80&0.010&$\bigcirc$& &-15.22& &-14.20&2.29&1.03\\
0.85&0.010&$\bigcirc$& &-16.59& &-15.84&1.63&0.75\\
0.90&0.010&$\bigcirc$& &-17.81& &-17.56&1.33&0.25\\
1.00&0.010&$\bigcirc$& &-18.97& &-18.86&1.09&0.12\\
1.10&0.010&$\bigcirc$& &-19.51& &-19.42&0.91&0.09\\
1.20&0.010&$\bigcirc$& &-19.83& &-19.76&0.72&0.07\\
\hline
1.00&0.020&$\bigcirc$&-15.34&-19.00&-15.26&-18.96&1.22&0.04\\
0.85&0.025&$\bigcirc$&-14.13&-17.01&-13.42&-14.81&2.01&2.20\\
1.00&0.030&$\bigcirc$&-15.95&-19.06&-15.90&-19.02&1.22&0.04\\
1.00&0.040&$\bigcirc$&-16.40&-19.13&-16.31&-19.09&1.22&0.04\\
0.95&0.045&$\bigcirc$&-16.61&-18.88&-16.54&-18.18&0.98&0.70\\
\hline
0.60&0.050&$\bigcirc$& &-7.01& &-4.48&2.75&2.53\\
0.70&0.050&$\bigcirc$&-14.65&-11.96&-14.07&-9.36&4.46&2.60\\
0.80&0.050&$\bigcirc$&-9.75&-16.17&-9.87&-13.70&4.27&2.47\\
0.90&0.050&$\bigcirc$&-16.31&-18.54&-16.34&-17.91&1.29&0.63\\
1.00&0.050&$\bigcirc$&-16.05&-19.00&-15.61&-18.73&1.10&0.27\\
1.00&0.050&$\bigcirc$&-15.70&-19.00&-15.19&-18.73&1.07&0.28\\
1.00&0.050&$\bigcirc$&-16.66&-19.22&-16.57&-19.18&1.26&0.05\\
1.10&0.050&$\bigcirc$&-16.96&-19.68&-16.93&-19.61&0.90&0.06\\
1.10&0.050&$\bigcirc$&-17.08&-19.68&-17.11&-19.63&0.83&0.05\\
1.10&0.050&$\bigstar$&-11.27&-19.73&-11.51&-19.66&0.82&0.06\\
1.20&0.050&$\bigcirc$&-9.57&-19.93&-9.63&-19.87&0.58&0.06\\
1.20&0.050&$\bigstar$&-9.96&-19.97&-10.05&-19.89&0.65&0.08\\
\hline
0.95&0.055&$\bigcirc$&-16.70&-18.89&-16.65&-18.30&1.01&0.59\\
1.00&0.060&$\bigcirc$&-17.06&-18.81&-16.93&-18.87&1.08&-0.06\\
1.00&0.070&$\bigcirc$&-17.24&-19.32&-17.08&-19.29&1.18&0.03\\
\hline
0.60&0.080&$\bigcirc$&-13.55&-11.98&-12.36&-10.14&6.55&1.85\\
0.70&0.080&$\bigcirc$&-16.43&-14.43&-16.42&-12.14&6.14&2.29\\
0.80&0.080&$\bigcirc$&-16.93&-17.06&-16.75&-14.47&3.65&2.59\\
0.90&0.080&$\bigcirc$&-16.95&-18.81&-16.93&-18.06&1.12&0.75\\
0.90&0.080&$\bigstar$&-15.95&-19.01&-15.92&-18.95&1.04&0.06\\
1.00&0.080&$\bigcirc$&-17.49&-19.38&-17.42&-19.35&1.20&0.03\\
1.00&0.080&$\bigstar$&-11.11&-19.51&-11.34&-19.48&0.94&0.03\\
1.10&0.080&$\bigcirc$&-17.31&-19.79&-17.34&-19.73&0.79&0.06\\
1.10&0.080&$\bigstar$&-8.80&-19.84&-8.56&-19.79&0.76&0.06\\
1.20&0.080&$\bigstar$&-8.79&-20.01&-6.81&-19.96&0.55&0.05\\
\hline
1.00&0.090&$\bigcirc$&-17.58&-19.41&-17.54&-19.39&1.16&0.02\\
0.70&0.100&$\bigcirc$&-16.99&-15.12&-16.85&-12.29&5.86&2.83\\
1.00&0.100&$\bigcirc$&-17.73&-18.86&-17.73&-18.95&1.13&-0.10\\
1.10&0.100&$\bigcirc$&-8.56&-19.89&-7.68&-19.84&0.68&0.05\\
1.20&0.100&$\bigcirc$&-9.20&-20.01&-8.61&-19.98&0.45&0.03\\
\enddata
\tablecomments{ A $\bigcirc$ indicates a double detonation while a $\bigstar$ indicates an edge-lit ignition.}
\end{deluxetable}

While qualitatively similar to the results seen in the double detonation models presented in \citet{2017MNRAS.472.2787N}, the $g-r$ colors of our models are noticeably redder than theirs, likely due to differences in how the radiation transport is handled in STELLA \citep{2000ApJ...532.1132B} compared to Sedona.

\citet{2017Natur.550...80J} present evidence for a SN~Ia trigger by a helium flash. The characteristics of SN~2016jhr are a kink in the early lightcurve, coupled with a color which evolves from the blue to the red. This is similar to the behavior of many of our models (see Figure~\ref{fig:musses}), though in our parameter sweep we do not find a perfect match to this SN~Ia. We do note that this SN~Ia is similar to the very rare class of SNe~Ia which include SN~2006bt \citep{2006bt} and PTF10ops \citep{PTF10ops} - SNe~Ia which have broad lightcurves that are incongruent with their peak brightness with respect to the Phillips relationship \citep{phillips1993}.

\section{Spectra}
\label{sec:spectra}
This section examines the synthetic spectra produced for our models with Sedona. Again, models are grouped into two categories: thick shell models and thin shell models. The former exhibit peculiar features due to the radioactive material in their outermost ejecta that serve to distinguish them from a normal SN~Ia. The later reproduce several typical SN~Ia signatures.

\subsection{Thick shell models: predictions for future searches}

\begin{figure}[tbph]
  \centering
  \includegraphics[width=1\columnwidth]{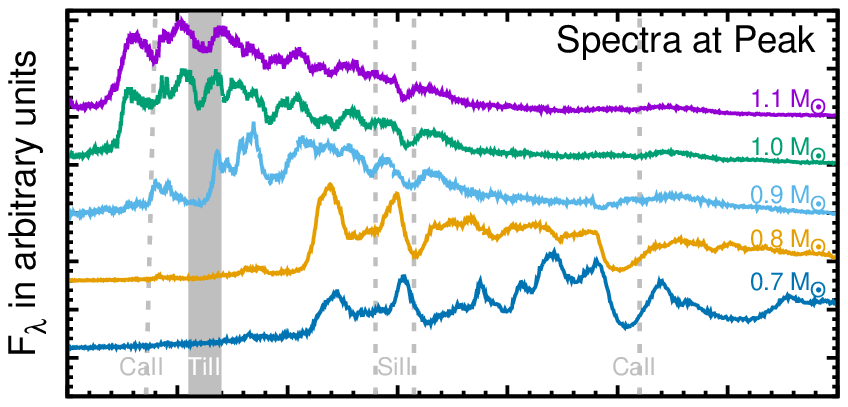}\vspace{-3em}\\
  \includegraphics[width=1\columnwidth]{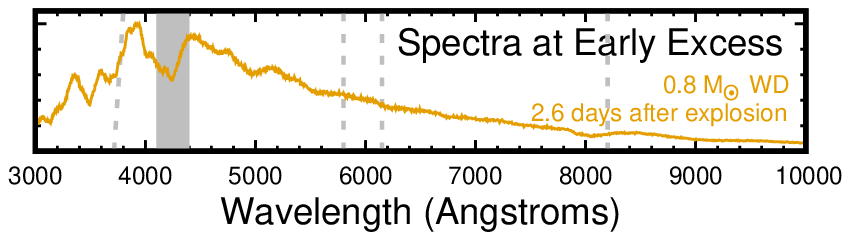}
  \caption{Spectra of the double detonations with thick (0.08 M$_\sun$) helium shells at peak in the top panel and an example spectrum during an early excess flux peak in the bottom panel, at which time for all models the spectra look qualitatively the same. The spectra of the thick shell models exhibit significant line blanketing at low wavelengths due to the ashes of the burnt helium shell. Models with 0.08 M$_\sun$  helium on WDs less than 0.8 M$_\sun$  are completely line blanketed for wavelengths less than 5200 \AA. The spectra taken at early excess is almost featureless with the exception of a \ion{Ti}{2} trough present from 4000-4300 \AA. These predicted signatures have yet to be seen in an observed SN~Ia.} 
  \label{fig:thickshellspectra}
\end{figure}

The top panel of Figure~\ref{fig:thickshellspectra} shows the spectra at g-band peak for several models with 0.08 M$_\sun$  helium shells. The spectra at g-band peak exhibit significant line blanketing at low wavelengths due to the ashes of the burnt helium shell. Models with 0.08 M$_\sun$  helium on WDs less than 0.8 M$_\sun$  are completely line blanketed for wavelengths less than 5200 \AA. We predict this spectral signature, that has yet to be seen in an observed SN~Ia, even though models with a total mass more than 0.9 M$_\sun$ are bright enough to be easily discovered at peak (see Figure~\ref{fig:gbandshells}) for many current surveys. As such, if events like these exist in nature, they must be intrinsically rare.

The bottom panel of Figure~\ref{fig:thickshellspectra} shows an example spectrum taken at the time of the early flux excess due to radioactive decay of material created from the burnt helium shell (typically 2-3 days after explosion). These early spectra all look qualitatively similar: a blue continuum which is mostly featureless above ~4500 \AA\ while bluer wavelengths show a \ion{Ti}{2} trough ($\sim$4000-4300 \AA) and a \ion{Ca}{2} H\&K absorption feature ($\sim$3700 \AA). Because spectra at this early flux excess all look similar only one example (from the 0.8 M$_\sun$  WD) is included in the plot. 

We note that 3-D simulations of deflagration driven Chandrasekhar mass CO WD explosions have also produced $^{56}$Ni mixed into the outer layers of SN ejecta \citep{Fink2014}. While these explosions also produce red, subluminous, lightcurves they lack the early flux excess and extreme UV line blanketing seen in our thick shell spectra and thus would be distinguishable from a sub-Chandrasekhar explosion.

\subsection{Thin shell models: relationship to observed SNe~Ia}

\begin{figure}[tbph]
  \centering
  \includegraphics[width=1\columnwidth]{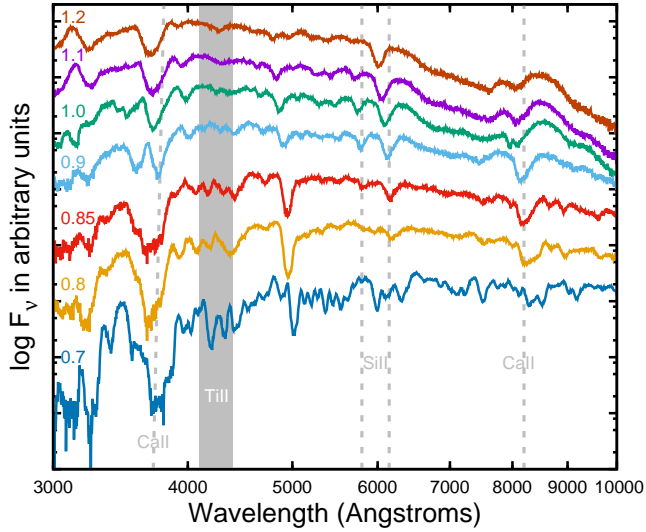}
  \caption{A peak-brightness spectral series similar to that in \citet{Nugent95ApJ}, following just the supernovae formed through the double detonation process with helium shells of 0.01 M$_\sun$. The velocity of the \ion{Si}{2} line at 6150~\AA\ points to an overall mass-luminosity-velocity relationship as the line velocity increases for the more massive WDs.} 
  \label{fig:nugent_si2}
\end{figure}

The very thin shell models in this survey, those with 0.01 M$_\sun$  helium on their surface, exhibit similar characteristics to many normal SNe~Ia. In Figure~\ref{fig:nugent_si2} we present the spectra of these models at g-band peak ordered by total mass. 
The thin shell models exhibit many of the characteristics expected from the spectroscopic series in \citet{Nugent95ApJ}. The models show the characteristic change in the ratio of the \ion{Si}{2} lines. The 5800 \AA\ feature decreases in strength with respect to the 6150 \AA\ feature, as the luminosity (and mass) of the models increase. As the luminosity (mass) of the models increase, we also see \ion{Fe}{2} transitioning to \ion{Fe}{3}. Lastly, we see \ion{Ti}{2} features begin to appear in low luminosity (mass) models. 

\begin{figure}[tbph]
  \centering
  \includegraphics[width=\columnwidth]{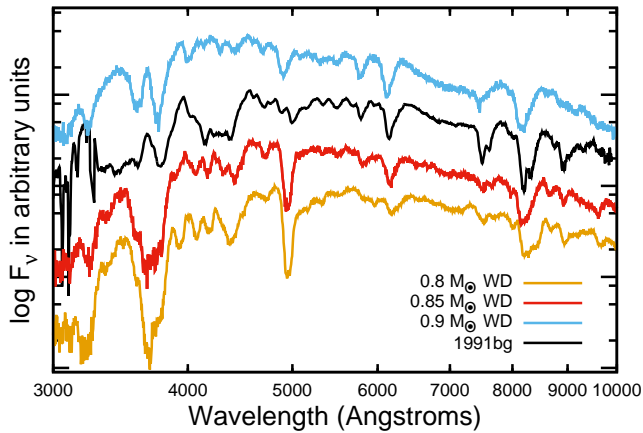}
  \caption{A comparison of SN~1991bg to the 0.8, 0.85 and 0.90 M$_\sun$  WDs with helium shells of 0.01 M$_\sun$. The number of SNe~Ia like SN~1991bg \citep{1991bg, 2011MNRAS.412.1473L} and the sensitivity of our models to total mass imply that either there is some natural floor to exploding SNe~Ia at this mass ($M_{tot} \sim0.85$ M$_\sun$) or that these SNe are created by another mechanism.} 
  \label{fig:91bg}
\end{figure}

As noted by \citet{Shen2018} and seen in Figure~\ref{fig:91bg}, the thin shell WDs with mass 0.85 M$_\sun$  are able to reproduce the features seen in the subluminous SN~1991bg class of objects. Interestingly, just below this mass the features do not resemble those of any known supernovae.  

Given the search conducted by \citet{2011MNRAS.412.1473L} and the sheer number of 1991bg-like SNe~Ia they discovered, we believe that they would have been sensitive to finding SNe~Ia slightly below this mass given that their luminosity cut-off was 1-2 magnitudes fainter in their search than the peak brightness for the 1991bg-like objects ($M_B\sim-17.25$). Coupled with the mass function for CO WDs increasing below this mass \citep{2018MNRAS.480.3942H}, it is thus likely that the 0.85 M$_\sun$  WDs represent some minimal total mass in the ignition of sub-Chandrasekhar mass explosions, or that there is an entirely separate mechanism for these supernovae. 

\subsection{Thin shell models: implications of kinetic energy --- mass relationship}

One of the interesting properties of the sub-Chandra-sekhar class of models is that there is an implicit relationship between the mass of the WD and the total $^{56}$Ni created. For the models which resemble observed SNe~Ia (0.85 M$_\sun$  $<$ $M_{WD}$ $<$ 1.2 M$_\sun$), the gravitational binding energy only increases by a factor of 3.5 while the $^{56}$Ni mass increases by two orders of magnitude over this range in mass. Consequently this further implies a direct relationship between the mass and the kinetic energy, as the kinetic energy in these explosions is a direct result of the CO synthesized to heavier elements in the explosion minus the gravitational binding energy of the WD. This can be seen in Figure~\ref{fig:nugent_si2} and in Figure~15 of \citet{Shen2018} where the \ion{Si}{2} velocity is proportional to the mass of the WD. While we have only explored WD with 50-50 carbon-oxygen fractions, \citet{Shen2018} shows that when one explodes WDs with ratios of 30-70, the velocity drops by a few hundred km/s and the luminosities decrease by $\sim10$\%. Thus, even spanning the full possible range in the permissible ratio of carbon to oxygen, there will exist a very tight relationship between luminosity and the \ion{Si}{2} velocity. 

\begin{figure}
    \centering
    \includegraphics[width=\columnwidth]{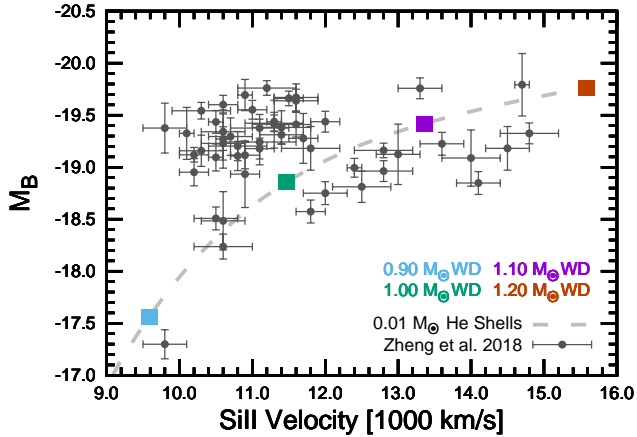}
    \caption{A plot of the \ion{Si}{2} velocity vs. $M_B$ (both at B-band peak) for those SNe~Ia found in \citet{2018ApJ...858..104Z} and our thin shelled models which are connected by a spline fit (dashed line) to guide the eye. We note there is a cluster of SNe~Ia that follow this relationship and a tight bunch that are offset and at lower velocities, given their luminosity, from our models.} 
    \label{fig:SiMB}
\end{figure}

In Figure~\ref{fig:SiMB} we explore this further by plotting our models on top of the extinction corrected observed SNe~Ia found in \citet{2018ApJ...858..104Z} in a plot of $M_B$ vs. the velocity of \ion{Si}{2}. What becomes immediately apparent given the location of our models, is that there appears to be two separate clusters of SNe~Ia. The first cluster traces our models and follows the relationship between brightness and velocity that is a natural byproduct of the sub-Chandrasekhar class of explosions. The second cluster is offset from this relationship, quite tightly bunched, and is considerably slower given their median luminosity than our models.

This second cluster likely represents explosions at the Chandrasekhar limit. The difference in velocity of this clump from our models, at the same luminosity, is what one would expect if these were near the Chandrasekhar limit. This is a combination of the effect of more graviational binding energy for these WD as well as the differences in the underlying explosion mechanism. Photometric evidence has already pointed to the need for sub-Chandrasekhar mass SNe~Ia. \citet{2014MNRAS.445.2535S} show empirical fits to the lightcurves of the SNLS and SDSS SNe~Ia which necessitate 30\% of these events originate from sub-Chandrasekhar progenitors. Furthermore, \citet{2018ApJ...852L..33G} show that lightcurve fits to low-luminosity events require a population of sub-Chandrasekhar explosion mechanisms. Here, we demonstrate, for the first time, both photometric {\it and} spectroscopic evidence to identify a sub-class of sub-Chandrasekhar SNe~Ia which span the range of normal to subluminous SNe~Ia.

\begin{figure}[pb!]
    \centering
    \includegraphics[width=\columnwidth]{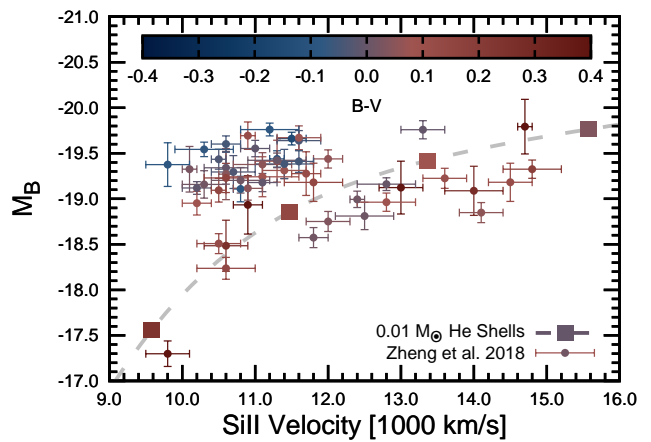}
    \caption{A plot of the \ion{Si}{2} velocity vs. $M_B$ (both at B-band peak) for the models and the observerd SNe~Ia of \citet{2018ApJ...858..104Z}. The color scale shows the $B_{max}-V_{max}$ colors for each SN~Ia, though for the observed SNe~Ia we have only corrected for MW extinction. Note that the models are significantly redder than the colors seen for those SNe~Ia found in the clump to the upper left, while the SNe~Ia that follow the monotonic relationship between velocity and brightness have similar colors to the models. }
    \label{fig:SiMV}
\end{figure}

This is not the first time that these supernovae have been called out as being different. \citet{2009ApJ...699L.139W} separates out observed SNe~Ia into ``normal" and ``high-velocity" groupings, split at 11,800 km/s, arguing that these clusters have different dust environments and thus their extinction corrections should be different. \citet{2011ApJ...729...55F} argued that the intrinsic colors for these two groups are different, that there is a tight correlation between velocity and color, and that these two groups are offset from each other in color space. In Figure~\ref{fig:SiMV} we highlight the color properties of these two clusters of supernovae and our models.  It should be noted that while the majority of SNe~Ia with peak \ion{Si}{2} velocities in excess of 11,800 km/s follow our sub-Chandrasekhar models, there is a significant population of SNe~Ia at lower velocity (as well as the 1991bg-like SNe~Ia) which do as well. Thus making a single cut at one velocity insufficient to delineate these two populations.

What is further indicative of two separate mechanisms is that some members of the putative clump of Chandrasekhar mass SNe~Ia show uncorrected colors bluer than any of our models, while the higher velocity SNe~Ia are similar to, or redder than, our models (which can be accounted for by normal extinction due to dust). This further supports the clustering of the observed SNe~Ia into these groups, though we note that we have employed a single extinction correction (following \citet{2018ApJ...858..104Z}) in the calculation of $M_B$ and more work here needs to be done to carefully sub-classify and identify individual SNe~Ia as being in one group or another. What this does imply is that corrections for extinction due to dust must take into account the intrinsic colors of the sub-classes of SNe~Ia, and that this can not be simply done with a single velocity cut to separate these groups. Furthermore it obviates the need for different $R_V$'s for the high-velocity group of SNe~Ia.   

Finally, we note a potential prediction of our models with respect to the division between these two clusters of SNe~Ia. In all of our models that match the observed SNe~Ia (the thin shell models with 0.85 M$_\sun$  $<$ $M_{WD}$ $<$ 1.2 M$_\sun$), the amount of unburnt carbon is quite minimal with only 0.001 M$_\sun$  in the 1991bg-like events dropping to 10$^{-5}$ M$_\sun$  or less for the more normal SNe~Ia. The efficiency of carbon-burning is a characteristic of the sub-Chandrasekhar models, though more work will have to be done with multi-dimensional simulations to verify this property. Even still, examining several 1-D Chandrasekhar-mass models shows that the amount of unburnt carbon can be quite substantial. The pure deflagration model W7 of \citet{1984ApJ...286..644N} and the pulsating delayed detonation models seen in \citet{1995ApJ...444..831H} typically have $\sim$0.03 M$_\sun$  of carbon left over after explosion. 

Observationally there may be evidence for this as well. \citet{2014MNRAS.444.3258M} notes that carbon spectral features are not ubiquitous in well observed SNe~Ia, and that $\sim$40\% show no hint of \ion{C}{2} features in their spectra when observed at times earlier than −10 days before peak brightness. What is interesting here is that none of their high-velocity SNe~Ia sample show signs of carbon in the spectra. Further, \citet{2011ApJ...732...30P} show that most of the objects in their sample that exhibit the \ion{C}{2} 6580 \AA\ absorption features are of the low-velocity gradient sub-type. \citet{2011ApJ...743...27T} argue that the lack of carbon in high velocity SNe~Ia might be due to to line blending effects. However, our models would suggest that these are sub-Chandrasekhar mass explosions and that the lack of a \ion{C}{2} feature is due to an absence of carbon in their atmospheres. What is further intriguing is that both \citet{2014MNRAS.444.3258M} and \citet{2011ApJ...743...27T} note that those SNe~Ia with carbon features are grouped tightly in lightcurve shape space and have colors that are bluer than most SNe~Ia. Our sub-Chandrasekhar models exhibit neither of these characteristics. This highly motivates future work where all of these properties are examined simultaneously in well observed SNe~Ia.

\section{Discussion}
\label{sec:discussion}
We have presented the results and analysis of a survey of 1-D sub-Chandrasekhar mass WD explosion models. These models can be separated into two categories, those with thin and thick helium shells. 
The thick helium shell models are considerably different from the SNe~Ia observed to date, with the possible exception of the SN~2006bt and PTF10ops like supernovae. The following characteristics summarize their properties:

\begin{itemize}
    \item[--] A detectable early excess in flux in the SNe~Ia light curves during the first few days after explosion caused by radioactive material burned in the helium shell.
    \item[--] Significant blanketing in the UV and the blue portion of the optical spectra near peak due to the Fe-peak elements in the helium shell ashes.
    \item[--] The very early spectra show a handful of spectroscopic features in the near-UV through 5000\AA, but are otherwise featureless.
    \item[--] A distinct red-to-blue evolution of the optical colors shortly after explosion
\end{itemize}

The models with thin helium shells exhibit lightcurves and spectral features similar to normal SNe~Ia. We show an inherent correlation between mass, brightness and velocity of the spectral features exhibited by these models. Noting this, and comparing to a recent set of well-observed SNe~Ia, we see further evidence for at least two distinct populations of SNe~Ia either having Chandrasekhar-like masses or sub-Chandrasekhar masses. These groupings can be further separated by the presence or absence of carbon features in their spectra, as carbon is almost completely burned in our models, while unburnt carbon is present in most Chandrasekhar-mass simulations. 

The implications for using SNe~Ia as cosmological probes are two-fold. First, it is likely that additional lightcurve features and/or a spectrum will be needed to sub-classify these events. Second, as the intrinsic colors of these two populations are different, this will necessitate different extinction corrections for each sub-class. 

This work has laid the foundation for future 3-D studies of these explosions which may reveal distinct line-of-sight differences that could be probed by spectropolarimetry, allowing us to further disentangle the potential progenitor systems for SNe~Ia. Furthermore, the effects of even minimal mixing in the outermost ejecta during hydrodynamic evolution can lead to changes in the early flux excess present in our thick shell models. Future studies will probe how much mixing is realistically achievable through higher dimensional simulations, carefully examine nucleosynthetic yields from the burning helium shells, and place limits on the amount of unburnt carbon that remains after explosion.  

\acknowledgments{
We thank the anonymous referee for helpful comments that improved the clarity of this paper as well as comments and suggestions by Ken Shen. This work was supported by the National Science Foundation Graduate Research Fellowship Program under Grant No. DGE 1752814. D.N.K is supported in part by the U.S. Department of Energy, Office of Science, Office of Nuclear Physics, under contract number DE-AC02-05CH11231 and DE-SC0017616, and by a SciDAC award DE-SC0018297. This research used resources of the National Energy Research Scientific Computing Center, a DOE Office of Science User Facility supported by the Office of Science of the U.S. Department of Energy under Contract No. DE-AC02-05CH11231 as well as the savio cluster at UC Berkeley.  
}

\software{Sedona (Kasen et al. 2006), Castro (Almgren et al. 2010), MAESTRO (Zingale et al. 2018; Nonaka et al. 2010)}

\bibliographystyle{yahapj}
\bibliography{main.bib}


\end{document}